\documentclass[12pt]{article}
\usepackage{amsmath}
\usepackage{amsfonts}
\usepackage{amssymb}

\numberwithin{equation}{section}

\newcommand{\ns}{\normalsize} 
\newcommand{\cK}{\mathcal{K}}
\newcommand{\cM}{\mathcal{M}}
\newcommand{\cN}{\mathcal{N}}
\newcommand{\cP}{\mathcal{P}}
\newcommand{\cG}{\mathcal{G}}

\newcommand{\ox}{\omega}
\newcommand{\tox}{\tilde \omega}
\newcommand{\txi}{\tilde \xi}
\newcommand{\nn}{\nonumber}
\newcommand{\fx}{\delta}  
\newcommand{\Ah}{\alpha}  
\newcommand{\Bh}{\beta}   
\newcommand{\Y}{\hat{Y}}

\begin{document}

\begin{titlepage}

\title{
  {\ns \hfill{hep-th/0608171}\\
   \hfill{ZMP-HH/2006-12}\\}
    \vskip 2cm
   {\Large\bf Heterotic--type IIA duality with fluxes}\\[0.5cm]}

\author{{\ns Jan Louis$^{1,2}$\footnote{email:
   jan.louis@desy.de}~} 
{\ns and Andrei Micu$^3$\footnote{email: amicu@th.physik.uni-bonn.de}
   $^{,}$\footnote{On leave from IFIN-HH Bucharest.}} 
  \\[0.5cm]
   {\it\ns $^1$II. Institut f\"ur Theoretische Physik der Universit\"at
   Hamburg}\\ 
   {\ns Luruper Chaussee 149, D-22671, Hamburg, Germany}\\[0.5cm]
   {\it\ns $^2$Zentrum f\"ur Mathematische Physik, Universit\"at Hamburg}\\
   {\ns Bundesstrasse 55, D-20146 Hamburg} \\[0.5cm]
    {\it\ns $^3$Physikalisches Institut der Universit\"at Bonn}\\
   {\ns Nussallee 15, D-53115, Bonn}}
\date{}

\maketitle

\begin{abstract}
  In this paper we study a possible non-perturbative dual of the
heterotic string compactified on $K3\times T^2$ in the presence of
background fluxes. We show that type IIA string theory
compactified on manifolds with $SU(3)$ structure can account for a
subset of the possible heterotic fluxes. This extends our previous
analysis to a case of a  non-perturbative duality with fluxes.
\end{abstract}

\vfill August, 2006

\thispagestyle{empty}

\end{titlepage}

\section{Introduction}

Compactifications with fluxes have a received some attention recently
for a number of different reasons \cite{MG1}. For example, they are often
necessary in the construction of string backgrounds which include
non-trivial D-branes. At the same time it has been realized that they
generate a potential for (some of) the scalar moduli present in all
supersymmetric string compactifications. This in turn can lift part of
the vacuum degeneracy of a given string background and
lead to more realistic scenarios of particle phenomenology or cosmology.

A geometrical `cousin' of flux compactifications are string backgrounds with a
space-time which is of the form $M^{1,3}\times \Y$ where $\Y$ is a compact
six-dimensional manifold with $G$-structure. Such manifolds are natural
generalizations of Calabi-Yau and/or 
$G_2$ manifolds in that they also admit a globally defined spinor but this
spinor is not covariantly constant.  As a consequence a scalar potential is
induced and, exactly as for flux backgrounds, part of the vacuum degeneracy is
lifted \cite{MG1}.

One of the interesting questions concerning these new backgrounds is the fate
of perturbative and non-perturbative dualities which hold for
compactifications on Calabi-Yau manifolds.  For example, 
it has been shown in refs.\ 
\cite{GLMW}-\cite{Chuang:2005qd}  
that mirror
symmetry between type IIA and type IIB string theories compactified on mirror
manifolds can be maintained in the presence of background fluxes if the
compactification manifold is chosen to be within a specific class of manifolds
with $SU(3)$ structure.  
Similar considerations have been carried out for a variety of other
backgrounds including geometrical \cite{Becker:2004qh}--\cite{Dasgupta:2006yd}
and non-geometrical set-ups \cite{Dabholkar:2002sy}--\cite{Shelton:2006fd}.

Mirror symmetry is a perturbative duality in that it does not act on the
dilaton and holds at weak coupling of both mirror symmetric backgrounds.
However, it is of obvious interest to also study non-perturbative dualities in
the presence of background fluxes and/or for compactifications on manifolds
with $G$-structure.  In this case the dilaton is non-trivially involved in the
duality map and hence the analysis becomes more complicated.

In this paper we study an example of a `generalized'
non-perturbative duality. We consider a subset of possible
fluxes in compactifications of the heterotic string on $K3\times T^2$
and show that a candidate dual is the type IIA string compactified on
manifolds with $SU(3)$ structure. This generalizes the
non-perturbative duality between the heterotic string on 
$K3\times T^2$ and  type IIA strings compactified on Calabi-Yau
threefolds \cite{Kachru:1995wm}--\cite{AL}. A first
step in this direction was undertaken in ref.\ \cite{CKKL} where it
was shown that the dual of the Abelian gauge field strength fluxes through a
certain cycle on the heterotic side corresponds to turning on (electric) RR
fluxes on the type IIA side. In this paper  we turn on 
a different set of fluxes in the heterotic string and argue 
that their type IIA dual corresponds to the torsion
of a $SU(3)$-structure manifold considered previously in
refs.~\cite{AT,GLW,Ferrara}.
We perform the analysis at the
level of the $N=2$ effective action for a whole class of such
compactifications. 

The paper is organized as follows. In section~\ref{fluxold} we briefly recall
the necessary facts about the heterotic string compactified on $K3\times T^2$
with background fluxes following \cite{LM1}. In \ref{fluxnew} we slightly
generalize our previous analysis in that we choose a more general solution to
the Bianchi identity of the NS B-field. This in turn leads to a more general
form of the resulting potential. In section~\ref{typeIIA} we propose a
non-perturbative dual compactification which consists of the type IIA string
compactified on a specific class manifolds with $SU(3)$ structure.  We compute
the effective action and in particular the Killing vectors and the potential.
In section~\ref{duality} we argue that the two actions are
equivalent and section~\ref{conclusions} contains our conclusions. In order to
make the paper self-contained we include two appendices with some well known
facts about heterotic and type IIA compactifications with $N=2$ supersymmetry
in four dimensions. In both appendices we also display some of the more
technical details which are needed in order to show the consistency with $N=2$
gauged supergravity of the compactifications studied in the main text.

\section{The Heterotic string compactified on $T^2 \times K3$}
\label{heterotic}

In this section we briefly review the compactification of the heterotic string
on the six-dimensional manifold $T^2 \times K3$ with background fluxes 
following \cite{LM1}. However, we do not consider the most general
set of fluxes but instead focus only on 
fluxes for $U(1)$ gauge fields along non-trivial two-cycles
of the $K3$ manifold. The reason is that for this set of fluxes we are
able to identify a type IIA dual background. 
In \ref{fluxold} we recall the results of
\cite{LM1} while in \ref{fluxnew} we slightly generalize our previous
analysis by allowing for a more general solution of the B-field
Bianchi identity.

\subsection{The effective action}\label{fluxold}

The low energy limit of the ten-dimensional heterotic string is
described by $N=1$
supergravity coupled to $N=1$ super Yang-Mills theory with gauge group
$SO(32)$ or $E_8 \times E_8$. The massless bosonic fields of the
gravitational multiplet are the metric
$g_{MN}$, the antisymmetric tensor field $B_{MN}$ and the dilaton
$\varphi$, while the gauge fields $A_M^a$ are members of 
vector multiplets. The
index $a$ runs over the adjoint representation of $SO(32)$ or $E_8 \times
E_8$ but for our purpose the specific choice of the gauge group is not relevant
and therefore we discuss both cases simultaneously.

The absence of anomalies in the ten-dimensional supergravity requires
that the $B$-field participates in a Green-Schwarz mechanism 
which leads to a modification of its field strength $H$
by appropriately normalized Lorentz- and Yang-Mills Chern-Simons terms
\begin{equation}
  \label{delB}
  H = dB + \omega_{\rm L} - \omega_{\rm YM}\ .
\end{equation}
As a consequence the Bianchi identity reads
\begin{equation}
  \label{Bid}
  dH = tr R \wedge R - tr F\wedge F \ ,
\end{equation}
where $R$ is the Riemann curvature tensor interpreted as a two form with
values in the Lie algebra of the local Lorenz group while $F$ represent the
field strengths of the gauge fields.
The details of the compactification (e.g. the precise light
spectrum) depend on the way in which this condition is implemented.  Since we
are mainly concerned with the computation of the potential we do not need to
specify a particular solution of \eqref{Bid} here.  For our purposes it is
sufficient to assume that \eqref{Bid} is satisfied, for example, by using the
standard embedding.  Whichever solution we choose it will generically break
the gauge group to some subgroup $G$ of $SO(32)$, or $E_8 \times E_8$. For
concreteness, we consider turning on an instanton configuration, $F_{inst}$, on
$K3$ with instanton number $\int_{K3} F_{inst} \wedge F_{inst} = 24$, so that
this cancels the contribution from the second Chern class, $\int_{K3} R \wedge
R = 24$, in \eqref{Bid}.

The $K3$ factor in the compactification breaks half of the 16
supercharges of the ten-dimensional theory. As a consequence the
four-dimensional effective theory has 8 supercharges corresponding to
$N=2$ supersymmetry. The massless spectrum contains the gravitational
multiplet with metric, 
gravitinos and graviphoton as components.
In addition there can be $n_v$ vector multiplets each with a vector, two
gauginos and a complex scalar  and 
$n_h$ hypermultiplets which contain four real scalars and two
hyperinos. 

A $K3$ manifold has 58 geometric moduli which combine with  22
axions coming from the internal $B$-field to form 20 hypermultiplets.
The 80 scalars of these multiplets
span the quaternionic coset manifold 
\begin{equation}
  \label{Mhhet1}
  \mathcal{M}_H = \frac{SO(4,20)}{SO(4)\times SO(20)}\ .
\end{equation}
$ \mathcal{M}_H$ is a submanifold of the entire quaternionic manifold which 
is a lot more
complicated and also contains moduli arising from the gauge bundle.
These moduli will play no role in the following and we will most of
the time set them to an arbitrary background value.

The massless vector fields have two different origins: first of all
one has the gauge fields of the unbroken gauge group $G$ and in
addition there are the Abelian Kaluza-Klein vectors of $T^2$.
Due to the $T^2$ factor in the compactification, the Yang-Mills theory
always has a Coulomb branch where $G$ is broken to its maximal Abelian
subgroup. In the following we assume that we are at a generic
point in the moduli space and only consider $n_v$ Abelian vector
multiplets coupled to supergravity.
Their scalar superpartners are the dilaton, the axion, the
moduli of $T^2$ as well as the gauge fields on the internal torus
which are scalars from a four-dimensional point of view. Altogether
these fields span the coset space 
\begin{equation}
  \label{Mvhet1}
  \mathcal{M}_V = \frac{SU(1,1)}{U(1)} \otimes \frac{SO(2,n_v -1)}{SO(2)
  \times SO(n_v-1)} \;.
\end{equation}

Let us now discuss the fluxes which we consider in this paper. They
arise from the gauge fields on $K3$ and therefore must be 
disentangled from the instanton
contribution.\footnote{We thank Andr\'e Lukas for drawing our attention
  on this point.} 
As explained before, the precise instanton configuration $F_{inst}$ which we
choose in order to satisfy \eqref{Bid}, generically breaks the gauge group to
some subgroup $G$ of $SO(32)$ or $E_8 \times E_8$.
By going to the Coulomb branch, the gauge group is further broken to the
maximal Abelian subgroup of $G$. From this point of view the
non-Abelian nature of the ten-dimensional gauge group is only
relevant for solving the constraint \eqref{Bid} and once it is 
solved we can discard these gauge fields together with all other fields
which become massive in this process. 
Now the fluxes we turn on are precisely for the `left-over' Abelian gauge
fields on $K3$ which also include the Kaluza--Klein vectors of $T^2$.
More precisely, we assume
that the following integrals are non-trivial
\begin{equation}
  \label{flux}
 \int_{\gamma^{\Ah}} F^I_{flux} = m^{\Ah I} \  , \qquad \Ah=1, \ldots , 22 \; ,
\end{equation}
where ${\gamma^{\Ah}}$ denotes the 22 non-trivial two-cycles of $K3$ and 
the index $I=0,\ldots , n_v$ label the Abelian vector fields on the
Coulomb branch.\footnote{$I=0$ counts the graviphoton which
is an Abelian vector in the theory but resides in the gravitational
rather than in a vector multiplet.}
Instead of defining the flux parameters  $m^{I \Ah}$ via the integral
\eqref{flux} we can equally well expand $ F^I_{flux}$ in terms of
a (real) basis of harmonic two-forms $\omega^\Ah$ on $K3$ which are dual
to the cycles ${\gamma^{\Ah}}$. This amounts to $ F^I_{flux} = m^{I \Ah}
\omega_\Ah$.

The only
thing we have to take care of at this point is that the fluxes can also
contribute to the right hand side of the Bianchi identity
\eqref{Bid}. This contribution -- when integrated over $K3$ -- is given by
\cite{KM} 
\begin{equation}
  \label{dBid}
  \fx = \int_{K3} F^I_{flux} \wedge F^J_{flux} \eta_{IJ} = m^{\Ah I} m^{\Bh J}
  \rho_{\Ah \Bh} \eta_{IJ} \; ,
\end{equation}
where $\rho_{\Ah \Bh}$ denotes the $K3$ intersection matrix, $\rho_{\Ah \Bh}=
\int \omega_\Ah \wedge \omega_\Bh$, which has signature $(3,19)$, while
$\eta_{IJ}$ is the invariant tensor on the $SO(2,n_v-1)$
factor of the moduli space defined in \eqref{Mvhet1} which has signature $(2,
n_v - 1)$. Within the set-up we
have presented so far, $\delta$ has to vanish for consistency. The reason
being that at this point we have already assumed that $F_{inst}$ saturates the
constraint \eqref{Bid} and the fluxes should not spoil this solution. In the
next section we relax this constraint but for now we impose $\delta=0$.

Without background fluxes the low energy effective action for this
compactification corresponds to an ungauged $N=2$ supergravity as 
it has been computed in refs.~\cite{Ceresole:1995jg,dWKLL}.
Turning on background fluxes gauges some of the isometries of the
moduli space and  generates a potential. 
One assumes that the fluxes are turned on
adiabatically so that the light spectrum does not change and that both
the string scale and the Kaluza-Klein scale are well above the scale
set by the fluxes. This ensures the consistency of the
compactification.

The low energy effective action for this compactification  was derived
in \cite{LM1} and shown to have the general form of
an $N=2$ gauged supergravity. The bosonic terms are found to be
\begin{equation}
  \label{s4het}
   S = \int \Big[ \frac12 R * \mathbf{1} 
   - g_{i\bar\jmath} d x^i \wedge * d \bar{x}^{\bar\jmath} 
   - h_{uv} D q^u \wedge *D q^v
  + \frac{1}{4}\, \mathrm{Im} \cN_{IJ} F^I \wedge * F^J  
  + \frac{1}{4} \, Re \cN_{IJ} F^I \wedge F^J - V \Big]  \ ,
\end{equation}
where the $x^i, i= 1,\ldots, n_v$ denote the complex scalars in the
vector multiplets, whose precise definition is given in appendix
\ref{hetT2K3}, while $q^u$ are the $4n_h$ real scalars of the
hypermultiplets. The $F^I = dA^I$ are the Abelian gauge field strength
and $\cN_{IJ}$ is the $N=2$ gauge coupling matrix given explicitly in 
\eqref{Nhet}.
$h_{uv}$ is the quaternionic metric on the hypermultiplet moduli space
which in general is unknown. For the subset of the 80 $K3$ moduli it
is the metric on the coset $SO(4,20)/SO(4)\times SO(20)$.
Finally, $g_{i \bar \jmath} = \partial_{x^i}
\partial_{\bar x^{\bar \jmath}} K$ is the (special) K\"ahler metric of the
vector multiplet  moduli space with $K$ given in \eqref{Kpoth}.

If no fluxes are turned on, the covariant derivatives $D q^u$ coincide
with partial derivatives and the potential $V$ in \eqref{s4het}
vanishes.  Turning on the 
fluxes \eqref{flux} gauges the Peccei-Quinn isometries associated
with the scalars coming from the $B$ field on $K3$.  This is easy to
see from the fact that gauge invariance of $H$ requires the $B$-field
to transform non-trivially in order to obey \eqref{delB}. More
precisely, for an Abelian gauge transformation, $\delta A^I= d
\lambda^I$ one needs the compensating transformation $\delta B =
\lambda^I F^J \eta_{IJ}$.  Expanding the $B$-field along $K3$ yields
the 22 axionic scalars $b^\Ah$ from\footnote{We have introduced a hat,
  $\hat{}$, in order to distinguish the ten-dimensional $B$-field from the
  four dimensional one.} 
\begin{equation}
  \label{Bexp}
  \hat B = B + b^\Ah  \omega_\Ah  \; ,
\end{equation}
where $B$ is an antisymmetric tensor in the four-dimensional space-time.
For the case of non-trivial background flux \eqref{flux}, we see immediately 
that the $b^\Ah $ transform as $\delta b^\Ah  = m^{I\Ah } \lambda^J \eta_{IJ}$ 
and
therefore  the effective action has to include the covariant derivative 
\begin{equation}
  \label{Db}
  D b^\Ah  = d b^\Ah  - m^{\Ah}_{I} A^I \; ,
\end{equation}
where we have defined
\begin{equation}
  \label{lowm}
  m^\Ah_I = \eta_{IJ} m^{\Ah J} \; .
\end{equation}

Apart from the covariant derivatives also a potential $V$ is generated by the
fluxes. It arises from the kinetic term for the gauge fields and therefore has
the form \cite{LM1}
\begin{equation}
  \label{Vflux}
  V_{flux} = - \frac12  h_{\Ah \Bh}\, m^{\Ah}_{I} m^{\Bh}_{J}\, 
  \left({\rm Im} \cN^{-1} \right)^{IJ}\ , 
\end{equation}
where $h_{\Ah \Bh}$ is the restriction of the quaternionic metric $h_{uv}$
to the space spanned by the charged scalars. As we just argued these
are precisely the axionic scalars $b^\Ah $
 and thus their $\sigma$-model metric can be derived 
from a direct reduction of the kinetic term of the $B$-fields. 
This yields
\begin{equation}
  h_{\Ah \Bh} =  \frac{1}{4 v}\int_{K3} \omega_\Ah  \wedge * \omega_\Bh \ ,
\end{equation}
where $v$ denotes the volume of $K3$.

To summarize, the fluxes gauge the $N=2$ supergravity in that they
induce the covariant derivatives \eqref{Db} and the potential
\eqref{Vflux}. The consistency with $N=2$ gauged supergravity was
shown in \cite{LM1}. However, in that proof it was essential that
$\fx$ defined in \eqref{dBid} vanishes.  It is this last constraint
which we now want to relax.

\subsection{Generalized fluxes}\label{fluxnew}

So far we have reviewed the results obtained in \cite{LM1} for
heterotic compactifications with fluxes. In particular, we chose the
fluxes to obey $\delta=0$ where $\delta$ was defined in \eqref{dBid}.
In this section we slightly generalize the setup in that we also
consider fluxes for which $\delta\neq0$ holds. The important point to
note is that requiring $\fx$ in \eqref{dBid} to vanish is quite an
arbitrary choice of having the Bianchi identity \eqref{Bid} satisfied.
The full integrated Bianchi identity in fact reads
\begin{equation}
  \label{fluxBid}
  0 = \int_{K3} \big[tr (F_{inst} \wedge F_{inst} ) - tr (R \wedge R) \big] +
  \fx \; .
\end{equation}
Demanding that both the integral as well as $\fx$ vanish separately is
only a special solution. Generically $\fx$ can be arbitrary as long as
we choose $F_{inst}$ in such a way that the above equation is
satisfied.  Let us take for now this point of view and discuss fluxes
with arbitrary $\fx$ and assume that the integral in \eqref{fluxBid}
is appropriately chosen such that this condition is satisfied.

We see immediately that the covariant derivatives
given in \eqref{Db} do not depend on the value of $\fx$
and therefore remain unchanged.  However, the potential \eqref{Vflux}
does change. Additional terms can arise from higher derivative 
($\alpha'$) corrections including  $(R^2)$-terms in the
ten-dimensional action. 
For a background which is Ricci flat -- as it is our case -- the
only non-vanishing terms are contractions of the Riemann curvature
tensor with itself. This combines with the kinetic term for the gauge fields
into
\begin{equation}
  \label{R2}
  S_{FR} = - \frac{1}{2} \int e^{- \varphi} \Big[ tr F \wedge *
  F - tr R \wedge * R \Big] \; ,
\end{equation}
where we have set $\alpha'=1$.
Let us now compute the contribution of the above term
to the potential in an instantonic background which satisfies \eqref{fluxBid}
for some arbitrary $\fx$.  For this we note that in order to preserve
supersymmetry $F_{inst}$ has to be of complex type $(1,1)$ and primitive
\cite{GSW}.  For a general $(1,1)$ form on a two-dimensional complex manifold
we can write
\begin{equation}
  \label{Fdual}
  (*F)_{a \bar b} = \epsilon_{a \bar b}{}^{c \bar d} F_{c \bar d} = 
  \epsilon_{ad} \epsilon_{\bar b}{}^c F_c{}^d = \frac12(g_{a \bar b} \delta_d^c
  - g_{d \bar b} \delta_a^c) F_c{}^d = \frac12 g_{a \bar b} F_c^c - \frac12
  F_{a \bar b} \; ,
\end{equation}
where $g_{a \bar b}$ is the metric on $K3$ written in complex coordinates and
we have used the decomposition of the four-dimensional $\epsilon$-symbol 
under complex indices $\epsilon_{ab}{}^{cd} = \epsilon_{ab} \epsilon^{cd} =
\delta_{ab}^{cd}$.  The primitivity condition for $F_{inst}$ further implies
that $(F_{inst})_a{}^a =0$ which inserted in \eqref{Fdual} implies
\begin{equation}
  \label{*F}
  *F_{inst} = -\frac12 F_{inst} \; . 
\end{equation}
The same is true for the curvature two-form. $K3$ is a K\"ahler manifold and
thus, $R$ is of type $(1,1)$. In addition, $K3$ is Ricci flat, which implies
the primitivity condition. Therefore, the contribution of the integral
\eqref{R2} to the potential, in an instantonic background satisfying
\eqref{fluxBid}, is
\begin{eqnarray}
  \label{Vinst}
  V_{inst} & = & \frac{1}{2} \int_{K3} e^{- \varphi} \Big[ tr F_{inst}
  \wedge * F_{inst} - tr R \wedge *R \Big] \nn \\
  & = & - \frac{1}{4} \int_{K3}
  e^{- \varphi} \Big[ tr F_{inst} \wedge F_{inst} - tr R \wedge R \Big] =
  \frac{1}{4} e^{- \varphi} \fx \; , 
\end{eqnarray}
where we have assumed that the dilaton is constant on $K3$.
We see that for a setup where the fluxes do not contribute to the Bianchi
identity, i.e. $\fx=0$, the instanton contribution to the potential precisely
cancels the contribution from the term $R^2$ due to the curvature of the
internal manifold. On the other hand, if we turn on fluxes such that $\fx\ne
0$, the instantons are no longer balanced against the internal curvature and
therefore the above term has to be taken into account in the potential. 
Thus, the complete potential, including also the flux contribution
\eqref{Vflux}, reads 
\begin{equation}
  \label{Vhet}
  V= V_{flux} + V_{inst} = -\frac12\, h_{\Ah \Bh}\, m^{\Ah}_{I} m^{\Bh}_{J}\,
  (\mathrm{Im} \cN^{-1})^{IJ} + \frac{e^{\phi}}{4 v}\,   m^{\Ah I} m^{\Bh J}
  \rho_{\Ah \Bh}\, \eta_{IJ} \; ,
\end{equation}
where we have multiplied \eqref{Vinst} by a factor $e^{2\phi}$ which
corresponds to a transformation into the Einstein frame and we have also
inserted the definition \eqref{dBid} of $\fx$. In the derivation above we
have used the fact that the ten-dimensional dilaton $e^{- \varphi}$ is
multiplied by the volume $V_6$ of the internal manifold in order to obtain the
properly normalized four-dimensional dilaton $e^{-\phi}=e^{- \varphi} V_6$.
We further assumed that the integration over the two-torus can be trivially
performed yielding a volume factor of $T^2$ which enters correctly in the
definition of the four-dimensional dilaton. On the other hand, the topological
$K3$ integral \eqref{fluxBid} does not yield a volume factor, $v$, of $K3$,
which therefore appears explicitly in \eqref{Vhet}.

The next task is to show that the potential $V$ of \eqref{Vhet} together
with the covariant derivatives \eqref{Db} is consistent with $N=2$
gauged supergravity. This check is presented in appendix \ref{gshet}.
Instead we now turn to our proposal for a dual type IIA background.

\section{Type IIA compactified on manifolds with $SU(3)$ structure}
\label{typeIIA}

The goal of this section is to identify a dual type IIA background.
Without fluxes heterotic compactifications on $K3\times T^2$ are 
non-perturbatively dual to type IIA compactified on Calabi-Yau
threefolds. This duality is non-perturbative in the sense that the
dilatons in both 
backgrounds are mapped to geometrical moduli and thus are not
constrained to be at weak coupling.\footnote{We briefly review this
  duality in section~\ref{duality}.}

A first obvious attempt is to turn on fluxes also on the type IIA side.
Indeed in ref.~\cite{CKKL} it was shown that RR fluxes on
the type IIA side which charge the axion under all vector multiplets,
correspond to gauge field fluxes on the heterotic side through the
$\mathbf{P}^1$ base of an appropriately fibered $K3$. However, the fluxes
through the other 21 two-cycles in $K3$ do not have obvious duals in the type
IIA picture.

Here we are going to propose that the dual of the heterotic fluxes can arise 
by modifying the compactification manifold. A similar approach has already been
pursued in the context of mirror symmetry.
In ref.~\cite{GLMW} it was shown that a certain class of manifolds
with $SU(3)$ structure -- termed half-flat manifolds -- are possible 
mirror duals to Calabi--Yau
compactifications of type IIB with NS three-form flux. 
These manifolds are characterized by the existence
of a globally defined and nowhere vanishing spinor $\eta$ which reduces the
structure group from $SO(6)$ to $SU(3)$. 
However, unlike in the Calabi-Yau case, this spinor is not covariantly
constant with respect to the Levi--Civita connection, but only with
respect to a connection with torsion. Or in other words the
(intrinsic) torsion measures the deviation of this spinor from being
covariantly constant. 

The existence of the spinor implies the existence 
of a $(1,1)$-form, $J$, and a $(3,0)$ form, $\Omega$, which are built
from appropriate spinor bilinears. As a consequence of $\eta$ being not
covariantly constant, both $J$ and $\Omega$ are not closed.
Instead they obey
\begin{equation}
  \label{dJO}
  \begin{aligned}
    (dJ)_{mnp} = & \; - 6 T_{[mn}{}^q J_{p]q} \; ,\\
    (d \Omega)_{mnpq} = & \; 12 T_{[mn}{}^r \Omega_{pq]r} \; ,
  \end{aligned}
\end{equation}
where $T$ denotes the intrinsic torsion. In this language Calabi-Yau manifolds 
are manifolds with $SU(3)$ structure for which the torsion vanishes
and hence $\eta$ is covariantly constant with respect to the Levi-Civita
connection and both $J$ and $\Omega$ are closed.

The existence of the spinor also ensures that  the low energy effective
action has $N=2$ supersymmetry.
The presence of torsion
gauges this supergravity and induces a scalar potential. Hence such
compactifications are natural candidates for duals of flux
compactifications which, as we saw in the previous section,  have exactly  
the same effect.

Type II compactifications on half-flat manifolds were studied in
refs.~\cite{GLMW,GM}, while ref.~\cite{GLW} considered a more general
class of $SU(3)$ structure compactifications. It is within this
generalized class of manifolds that we will locate the duals of
heterotic flux compactifications.

Since compactifications on manifolds with $SU(3)$ structure have
already been spelled out in some detail in
refs.~\cite{GLMW,GM,GLW,HP} we will be brief in the 
following and only concentrate on the important points.

The next step is to derive the low energy effective action of type IIA
supergravity compactified on manifolds with $SU(3)$ structure $\Y$ as proposed
in \cite{AT,GLW,Ferrara}.  We start from (massless) type IIA supergravity in
ten dimensions whose bosonic degrees of freedom consist of the graviton, $\hat
g_{MN}$, an antisymmetric tensor field, $\hat B_2$, and the dilaton, $\hat
\phi$, in the NS-NS sector and a one-form, $\hat C_1$, and a three-form, $\hat
C_3$, in the RR sector. The ten-dimensional action is given by
\begin{eqnarray}
  \label{SA10}
  S & = & \int \,e^{-2\hat\phi} \big( \frac12 \hat R \ast {\bf 1} + 2
  d \hat\phi \wedge * d \hat\phi - \frac14 \hat H_3 \wedge * \hat H_3 \big)
  - \frac12  \, \big(\hat F_2 \wedge * \hat F_2  + \hat F_4 \wedge * \hat F_4
  \big)  \nn \\
  & & - \frac12\Big[\hat B_2 \wedge d\hat C_3 \wedge d\hat C_3 - (\hat B_2)^2
  \wedge d\hat C_3 \wedge d\hat C_1 + \frac13 (\hat B_2)^3\wedge d\hat C_1
  \wedge d \hat C_1\Big] \; ,
\end{eqnarray}
where 
\begin{equation}
  \hat  F_4 = d\hat C_3 - d\hat C_1 \wedge \hat B_2\ , \qquad \hat F_2 =
  d\hat C_1\ , \qquad \hat H_3 = d \hat B_2 \ .
\end{equation}
The action is invariant under the following three independent Abelian
gauge transformations 
\begin{eqnarray}
  \label{gtrIIA}
  \delta \hat C_1 &=&  d  \hat\theta \ ,\qquad 
  \delta \hat C_3 = d \hat \Lambda_2\ , \\
  \delta \hat B_2 &=& d  \hat\Theta_1 \ , 
\quad  \delta\hat C_3 = \hat C_1 \wedge d \hat \Theta_1   \ .\nn
\end{eqnarray}

In order to perform the compactification we follow the
strategy outlined in ref.~\cite{GLW}. The first step is to decompose all
ten-dimensional fields under $SO(1,3)\times SU(3) \subset SO(1,9)$.
This merely amounts to a rewriting of the ten-dimensional fields in a
background with a smaller Lorentz group. If one identifies the
$SO(1,3)$ factor with the Lorentz group of a four-dimensional
  space-time the resulting effective theory 
has $N=8$ supersymmetry. This is most easily seen from the
decomposition of the ${\bf 16}$ spinor representation of  $SO(1,9)$
under $SO(1,3)\times SU(3)$ which yields
\begin{equation}
  \label{16decomp}
  {\bf 16} \to ({\bf 2},{\bf 3}) \oplus ({\bf 2},{\bf 1}) \oplus ({\bf
    \bar 2},{\bf \bar 3}) \oplus ({\bf \bar 2},{\bf \bar 1})\ .
\end{equation}
Here the ${\bf 2}$ and ${\bf \bar 2}$ denote complex conjugate Weyl
spinors of  $SO(1,3)$ and thus we see that four supersymmetries result
from the ${\bf 16}$-dimensional spinor representation. Since type IIA
has two gravitinos in the ${\bf 16}$, the effective theory has $N=8$
supersymmetry. 
One way to reduce the supersymmetry is to project out all $SU(3)$
triplets ${\bf 3}$ and only keep the singlets ${\bf 1}$. This singlet
is precisely the invariant spinor $\eta$ mentioned previously.
As can be seen from \eqref{16decomp} such a truncation has only one
gravitino from each of the two  ${\bf 16}$ or in other words $N=2$
supersymmetry from a four-dimensional point of view. The truncation
has to be implemented on the entire spectrum and, as shown in
ref.~\cite{GLW}, the resulting spectrum can be arranged in $N=2$ multiplets.

The next step is to Kaluza-Klein expand the ten-dimensional fields in
terms of a set of two-forms
$\omega_i$ (with dual four-forms $\tilde\omega^i$)
and a set of three-forms $(\alpha_A, \beta^B)$
on $\Y$.\footnote{No fields arise from one-forms or five-forms
  since they are $SU(3)$ triplets and projected out.} 
These forms are not necessarily
harmonic but $(1,\omega_i,\tilde\omega^i,\epsilon_g)$ form a
non-degenerate symplectic basis on the space of all even  
forms ($\epsilon_g$ denotes the volume from) and $(\alpha_A, \beta^B)$
form a
non-degenerate symplectic basis on the space of all three-forms.
In other words they obey  \eqref{norm2} and \eqref{norm3} exactly as
their Calabi--Yau `cousins'.
The expansion of the ten-dimensional fields in this basis also
resembles the Calabi-Yau situation reviewed in appendix~\ref{IIACYgen}
and reads
\begin{eqnarray}
  \label{fexp}
  \hat B_2 & = & B_2 + b^i \omega_i \; , \nn \\
  \hat C_1 & = & A^0 \; , \\
  \hat C_3 & = & C_3 + A^i \wedge \omega_i +  \xi^A \alpha_A - \tilde \xi_A
  \beta^A \; . \nn
\end{eqnarray}
Here $b^i$, $\xi^A$ and $\txi_A$ are scalar fields, $B_2$ is
a two-form in four dimensions which is dual to an axion $a$, $A^0$ and $A^i$ are
vector fields and $C_3$ is a three-form in four dimensions which is not
dynamical and dual to a constant.\footnote{Later on it will be 
essential to properly
perform this dualization as it generates terms which contribute
to the potential.}
Similarly, we expand $J$ and $\Omega$  as
\begin{equation}
  \label{JOexp}
  \begin{aligned}
      J = \; v^i \omega_i \; ,  \qquad
      \Omega =  \; Z^A \alpha_A - \cG_A \beta^A \; ,
  \end{aligned}
\end{equation}
where $v^i$ represent the analog of the Calabi--Yau K\"ahler moduli and $z^a =
Z^A/Z^0$ are the analog of the complex structure moduli. 
In \cite{GLW} it was further shown that
$\cG_A$ is the derivative of a prepotential $\cG(Z)$ 
with respect to the projective coordinates $Z^A$.

Altogether these
fields combine into $h^{(1,1)}$ vector multiplets, consisting of the
bosonic components $(A^i, x^i=b^i+i v^i)$ and 
$h^{(2,1)}$ hypermultiplets featuring the scalars $(z^a, \xi^a,
\txi_a)$.\footnote{We are abusing the notation here in that we denote
the dimension of the set of two-forms by $h^{(1,1)}$ and the dimension
of the set of three-form by $2+2h^{(2,1)}$.}
In addition there is a tensor multiplet with components
$(B_2, \phi, \xi^0, \txi_0)$ and finally $A^0$ is the graviphoton which
sits in the $N=2$ gravitational multiplet.

The compactification now proceeds in analogy with Calabi-Yau compactifications
which we recall in appendix \ref{IIACYgen} with the difference that the forms
in which we expand the fields are no longer closed. The generic case has been
discussed in ref.~\cite{GLW} but here we are only interested in the subclass
of $SU(3)$ manifolds which are dual to the heterotic compactifications.  Thus
we consider manifolds which obey the following differential relations
\begin{eqnarray}
  \label{doda}
  d \omega_i & = & - q_i^A \alpha_A \; , \nn \\
  d \alpha_A & = & 0 \; , \quad d \beta^A = q^A_i \tilde \omega^i \; , \\
  d \tilde \omega^i &=& 0 \; , \nn
\end{eqnarray}
where $q_i^A$ is a constant $n_v\times n_h = h^{(1,1)}\times
(h^{(1,2)}+1)$ matrix.
Using \eqref{JOexp} this amounts to
$ dJ= - v^i q_i^A \alpha_A$ and $d\Omega = \cG_A q^A_i \tilde \omega^i$.
The motivation for the choice \eqref{doda} is that the $q^A_i$ carry
one index $A$ which labels the hypermultiplets, 
and one  index  $i$ which labels the vector multiplets. This is precisely the
behavior we found in the heterotic flux compactification of the previous
section.

The derivation of the effective action proceeds analogously to the
Calabi-Yau case and yields a gauged $N=2$ supergravity of the form
\eqref{s4het}. 
The kinetic terms have exactly the 
same form as for Calabi-Yau compactifications which we recall in
appendix~\ref{IIACYgen}. In particular,
$g_{ij}$ is the metric on the space of vector multiplet
scalars $x^i = b^i+iv^i$ which has the form \eqref{gKd}, while
$\cN_{IJ}$ are the gauge couplings which have the form \eqref{N} and
can be obtained from the prepotential \eqref{FIIA}. Moreover,  the
quaternionic metric, $h_{uv}$, has precisely the form \eqref{qkt}
with the matrix $\cM$ defined as in \eqref{M}.

The effect of the torsion,
manifested through the derivatives \eqref{doda}, is
to turn some of the ordinary derivatives into covariant  derivatives
and induce a potential.  The covariant derivatives can be read 
off easily from the gauged isometries. Consider the gauge transformation of
the three-form potential $\hat C_3$ in ten dimensions
\begin{equation}
  \label{gtr10}
  \delta \hat C_3 = d \Lambda_2 \; ,
\end{equation}
and expand the two-form gauge parameter in the two-forms $\omega_i$
\begin{equation}
  \Lambda_2 = \lambda^i \omega_i \; .
\end{equation}
Using the differential relations \eqref{doda} and inserting the gauge
transformation in the expansion \eqref{fexp} we obtain the following
transformation properties 
\begin{equation}
  \label{gaugings}
  \begin{aligned}
    \delta \xi^A = & - \lambda^i q_i^A \; , \qquad
\delta A^i = & \; d \lambda^i \; .
  \end{aligned}
\end{equation}
The corresponding covariant derivative which is invariant under this
transformation reads
\begin{equation}
  \label{cdint}
  D \xi^A = d \xi^A + q^A_i A^i \; .
\end{equation}
Of course, the same conclusion can be reached by computing the gauge
invariant field strength $\hat F_4$ using the expansion \eqref{fexp}
\begin{equation}
  \label{F4exp}
  \hat F_4 = d C_3 - d A^0 \wedge B_2 + (d A^i - d A^0 b^i) \wedge
  \omega_i + (d \xi^A + q^A_i A^i) \wedge \alpha_A - d \txi_A \wedge
  \beta^A -\txi_A q^A_i \tilde \omega^i \; .
\end{equation}

Having found the charged fields, let us turn to the quaternionic
metric \eqref{qkt}. Notice that in these kinetic terms, the charged
fields $\xi^A$ appear also without a derivative and in order to render
this term gauge invariant we need to introduce a further term which
contains the gauge fields $A^i$. However, this can be avoided if we
perform a field redefinition $a \to a - \xi^A \txi_A$ under which
\eqref{qkt} becomes 
\begin{eqnarray}
  \label{newh}
  h_{uv} Dq^u \wedge * Dq ^v & = & \frac14 (d\phi)^2 + g_{a \bar b} dz^a
  \wedge * d \bar z^b + \frac{e^{4\phi}}{4} \, \Big[da - 2\tilde\xi_A D
  \xi^A \Big]^2  \\
  & & - \frac{e^{2\phi}}{2}\left(\mathrm{Im} \cM^{-1} \right)^{AB} 
  \Big[ D\tilde\xi_A - \cM_{AC} D\xi^C \Big]
   \wedge * \Big[ D\tilde\xi_B - \bar \cM_{BD} D\xi^D \Big]  \, ,\nn
\end{eqnarray}
where the covariant derivatives for the fields $\xi^A$ are given in
\eqref{cdint}.

By comparing \eqref{Db} with \eqref{cdint} we see that the type IIA
graviphoton does not appear in \eqref{cdint} while it does in
\eqref{Db}. Or in other words the type IIA side lacks ($h^{(2,1)}+ 1$)
parameters which are duals to $m^\alpha_0$.  
{}From ref.~\cite{GLW} we know that
the NS-NS fluxes combine with the torsion parameters 
\eqref{doda} as the zeroth component $q_0^A$, while from ref.~\cite{LM2}
we learn that the NS-NS fluxes result in gauge charges with respect to
the graviphoton.  These observations suggest that we also need to turn
on half of the NS-NS fluxes in order to recover all the flux parameters of
the heterotic compactification. Thus we choose
\begin{equation}
  \label{Hflux}
  H_{flux} = - q_0^A \alpha_A \; .
\end{equation}
Note that the Bianchi identity, $dH=0$, does not impose any additional
constraint but is automatically satisfied due to \eqref{doda}.

The correct treatment of such a flux requires that we perform a field
redefinition in the ten-dimensional type IIA action \eqref{SA10} so that only
$H=dB$ appears in the action and not the bare two-form potential $B$. However
in this basis it is more difficult to read off the correct vector degrees of
freedom in four dimensions and the calculation becomes more involved.
Therefore in order not to over complicate the calculation we shall set the
fluxes \eqref{Hflux} to zero at the beginning and switch them back on at the
very end using the following observation. Due to the differential relations
\eqref{doda} a shift in the vacuum expectation value for the scalars $b^i$ in
\eqref{fexp} of the form $b^i\to b^i + \rho^i$ will generate an $H$-flux of
the form $H_{ind} = - \rho^i q^A_i \alpha_A$, which is precisely the flux
\eqref{Hflux} if we identify\footnote{Note that if the rank of the matrix
  $q^A_i$ is not at least $h^{(2,1)}+1$ we can not in general absorb all the
  fluxes $q^A_0$ into the shifts $\rho^i$. However, in deriving the effect of
  these shifts we do not use arguments related to the rank of $q^A_i$ and
  therefore we can consider that $rank(q^A_i) \ge h^{(2,1)} +1$ and the result
  will also apply to the case $rank(q^A_i) \le h^{(2,1)}$.}
\begin{equation}
  \label{q0}
  q^A_0 = \rho^i q^A_i\ .
\end{equation}

The important thing to note here is that the $b^i$-shift is not a symmetry of
the theory since, due to \eqref{doda}, $H$ shifts as specified above.  If the
$\omega_i$ were closed, the $b^i$-shift would be a symmetry of the action and
the $b^i$ would be true axions of the low energy theory. As $\omega_i$ are not
closed the $b^i$-shift has to be accompanied by changes in other fields.  It
can be easily seen from the gauge transformation \eqref{gtrIIA} that we need
to perform an additional transformation on the ten-dimensional field $\hat
C_3$ which is of the form $\delta \hat C_3 = \rho^i \omega_i \wedge A^0 =
q^A_0 A^0 \wedge \omega_i$.  Since $\omega_i$ obeys \eqref{doda}, the field
strength $\hat F_4$ changes accordingly by $\delta \hat F_4 = q^A_0 A^0 \wedge
\alpha_A $.  Compared to the expansion \eqref{F4exp} one immediately sees that
this amounts to a change in the covariant derivative \eqref{cdint} for the
fields $\xi^A$, which can now be written
\begin{equation}
  \label{cdxi}
  D \xi^A = d \xi^A + q^A_I A^I \; .
\end{equation}
Note that the index $I$ now runs over all the vector fields in the theory,
including the graviphoton, as argued before based on the results in
\cite{LM2}.

Let us turn to the computation of the potential which is generated
by the torsion \eqref{doda} in the absence of $H$-fluxes.  There will
be several contributions to the potential which we shall analyze
separately in the following.  First of all one notices that the
relations \eqref{doda} effectively induce fluxes for the field
strengths $\hat F_4$ and $\hat H_3$.  From the expansion \eqref{fexp}
we find 
\begin{eqnarray}
  \label{inducedflux}
  (\hat  F_4)_{ind} =  - \tilde \xi_A q^A_i \tilde \omega^i \; , \qquad
  (\hat  H_3)_{ind}  =  - b^i q^A_i \alpha_A \; .
\end{eqnarray}
Clearly, these induced fluxes will generate potential terms when inserted in
the corresponding kinetic terms of \eqref{SA10}. Additional contributions arise
from dualizing the three-form $C_3$ and from the fact
that manifolds with $SU(3)$ structure described by the
differential relations \eqref{doda} are in general not Ricci flat. Let
is analyze these contributions one by one.

\begin{itemize}
\item[(i)] Internal fluxes

  Inserting \eqref{inducedflux} into the kinetic terms
  for $\hat C_3$ and $\hat B_2$ of the type IIA action \eqref{SA10}
and after performing
  the integration over the internal manifold we find
  \begin{equation}
    \label{VF4}
    V_F =  \frac{e^{4\phi}}{8 \cK}\, \txi_A \txi_B q^A_i q^B_j
    g^{ij} \; , 
  \end{equation}
  and
  \begin{equation}
    \label{VH}
    V_H = - \frac{e^{2 \phi}}{4 \cK}\, \cM_{AC} \left({\rm Im}
    \cM^{-1} \right)^{CD} {\bar \cM}_{DB}\, q^A_i q^B_j  b^i b^j \; ,
  \end{equation}
  where the matrix $\cM$ is defined in complete analogy with
  Calabi-Yau threefolds by the integrals given in \eqref{M}.

\item[(ii)] Dualization of $C_3$ in four dimensions 

After collecting all terms containing $C_3$ one can follow the standard
procedures for dualizing a three-form in four dimensions (see e.g.\
\cite{LM2}). Due to the non-trivial couplings of $C_3$, the
  dualization yields several terms one of which contributes
  to the potential.\footnote{The dualization also introduces an
additional flux parameter --the constant dual to $C_3$ -- but
for our purposes here we can  set it to zero from the
  beginning.}
We find
  \begin{equation}
    \label{VC}
    V_{C_3} = \frac{e^{4 \phi}}{2 \cK}\, (q^A_i b^i \txi_A)^2 \; .
  \end{equation}

\item[(iii)] Internal scalar curvature 

  The Ricci scalar for the manifolds
  with $SU(3)$ structure which we consider here was computed in \cite{dCGLM}. 
  After integrating over
  the internal manifold one obtains
  \begin{equation}
    \label{VR}
    V_R = - \frac{e^{2 \phi}}{4 \cK}\, \cM_{AC} \left({\rm Im} \cM^{-1}
    \right)^{CD} 
    {\bar \cM}_{DB}\, q^A_i q^B_j  v^i v^j + \frac{e^{2 \phi}}{2 \cK}
    e^{K(z)}\, 
    \cG_A \bar \cG_B q^A_i q^B_j (g^{ij} - 4 v^i v^j) \; ,
  \end{equation}
where $\cG_A$ are defined in \eqref{JOexp}.

\end{itemize}

Putting the above contributions together we obtain the final form of the
potential
\begin{equation}
  \begin{aligned}
    \label{potIIA}
    V  =\ &V_F+V_H+V_{C_3}+ V_R\\
    =\ &- \frac{e^{2 \phi}}{4 \cK}\, \cM_{AC} \left({\rm Im} \cM^{-1}
    \right)^{CD} {\bar \cM}_{DB}\, q^A_i q^B_j ( v^i v^j + b^i b^j)  \\
    & + \frac{e^{4 \phi}}{2}\, \txi_A \txi_B q^A_i q^B_j \left(\frac{g^{ij}}{4
        \cK} + \frac{b^i b^j}{\cK} \right) + \frac{e^{2 \phi}}{2 \cK}\, \cG_A
    \bar \cG_B q^A_i q^B_j (g^{ij} - 4 v^i v^j) \; .
  \end{aligned}
\end{equation}
In order to make the comparison with the heterotic side and also the relation
to the $N=2$ supergravity form of the potential more transparent, let us
regroup the terms and write it as
\begin{equation}
  \label{Valt}
  \begin{aligned}
    V = & \; \left[ - \frac{e^{2 \phi}}{4} \big(\cM {\rm Im} \cM^{-1} \bar \cM
      \big)_{AB} + \frac{e^{4 \phi}}{2} \txi_A \txi_B \right] \left(
      \frac{g^{ij}}{4 \cK} + \frac{b^i b^j}{\cK} \right) q^A_i q^B_j \\
    & + \left[ \frac{e^{2 \phi}}{16 \cK} \big(\cM {\rm Im} \cM^{-1} \bar \cM
      \big)_{AB} + \frac{e^{2 \phi}}{2 \cK} e^{K(z)} \cG_A \bar \cG_B \right]
      \left( g^{ij} - 4 v^i v^j \right) q^A_i q^B_j \; .
  \end{aligned}
\end{equation}

Recall that the potential above was computed in the absence of
$H$-fluxes. However, their inclusion at the level of the potential is
straightforward. We have argued above that $H$-fluxes characterized by
the parameters $q^A_0$ in \eqref{q0} can be turned on by incorporating
the shift  $b^i\to b^i + \rho^i$ with $\rho^i$ obeying \eqref{q0}.
Clearly, under this transformation only the first term in the potential
\eqref{Valt} changes and we can write the potential in its final form 
\begin{equation}
  \label{VIIA}
    \begin{aligned}
    V = & \; - \left[ - \frac{e^{2 \phi}}{4} \big(\cM \mathrm{Im} \cM^{-1}
      \bar \cM \big)_{AB} + \frac{e^{4 \phi}}{2} \txi_A \txi_B \right] \left(
      \mathrm{Im} \cN^{-1} \right)^{IJ} q^A_I q^B_J \\
     & - e^{2 \phi} \left[ \frac{1}{2} \big(\cM \mathrm{Im} \cM^{-1} \bar \cM
      \big)_{AB} + 4 e^{K(z)} \cG_A \bar \cG_B \right]
      \left( \frac12 (\mathrm{Im} \cN^{-1})^{IJ} + 4 X^I \bar X^J \right) q^A_I
      q^B_J \; , 
  \end{aligned}
\end{equation}
where $\mathrm{Im} \cN^{-1}$ is the inverse of the matrix given in \eqref{N} which has
the form
\begin{equation}
  \label{imN-1}
  (\mathrm{Im} \cN)^{-1} = -\frac{1}{\cK} \left(
    \begin{array}{cc}
      1 & b^i \\
      b^j & \frac{g^{ij}}{4} + b^i b^j
    \end{array}
    \right) \; .
\end{equation}
Note that we have also rewritten the second line in \eqref{Valt} in
terms of $\mathrm{Im} \cN^{-1}$ in order to make 
the symplectic structure of the
potential manifest. In this form the comparison  with
gauged supergravity and with the heterotic potential will become more
transparent.

The final task is to establish the consistency of the potential
with the general form of $V$ in 
$N=2$ gauged supergravity as we did on the heterotic side in
appendix~\ref{gshet}. The analogous check for the type IIA potential
\eqref{VIIA} is  done in appendix~\ref{gsIIA}. In the next section we
instead turn to the comparison between the heterotic and the type IIA action.

\section{Duality of heterotic and type II compactification}
\label{duality}

In this section we want to argue that the conjectured non-perturbative duality
between the heterotic string compactified on $K3\times T^2$ and the type IIA
string compactified on $K3$-fibred Calabi-Yau threefolds,
\cite{Kachru:1995wm,FHSV,KLT,KLM,AL}, can be generalized to a duality between
the heterotic string compactified on $K3\times T^2$ in the presence of a
specific set of fluxes and the type IIA string compactified on a particular
subclass of manifolds with $SU(3)$ structure. We discuss this at the
level of the bosonic effective actions which we derived in the previous two
sections.  We already noted that turning on fluxes and `turning on torsion'
does not alter the kinetic terms but only induces covariant derivatives and a
potential. Since the duality is already `established' for the kinetic terms in
principle, we only need to compare the covariant derivatives and the
potential. However, in order to do so we need to first fix the duality map
which precisely has to be done for the kinetic terms.

The idea is that for `every' Calabi-Yau threefold $Y$ there is a whole
family of manifolds with $SU(3)$ structure $\Y_T$ which share the
same light spectrum and the same kinetic terms but differ in the
torsion $T$ and, therefore, in covariant derivatives and potential.
For the case at hand this is expressed by the choice of the parameters
$q_I^A$. 

The duality between the heterotic string compactified on $K3\times T^2$ and
the type IIA string compactified on $K3$-fibred Calabi-Yau threefolds has been
established mainly in the vector multiplet sector
\cite{Kachru:1995wm,FHSV,KLT,KLM,AL}.  It has been shown that for Calabi-Yau
threefolds which have the structure of a $K3$ fibred over a ${\bf P_1}$ base
the vector multiplet couplings can be matched with the heterotic vacuum in the
limit of a large ${\bf P_1}$. The volume of the ${\bf P_1}$ is identified with
the heterotic dilaton such that a large ${\bf P_1}$ corresponds to weak
coupling on the heterotic side. Here we assume that precisely the same
situation carries over to manifolds with $SU(3)$ structure and a similar
identification can be made. (Obviously it would be very interesting to show
this in more detail.) The only caveat is that on the heterotic side the
natural variables appearing in a Kaluza-Klein reduction correspond to a
parameterization where the prepotential $F$ does not exist.  As reviewed in
appendix~\ref{hetT2K3gen} a symplectic rotation is necessary in order to
transform the action into a form that can be compared to the type IIA side.
This symplectic rotation is a symmetry for vanishing fluxes but is broken once
fluxes are turned on. Therefore it is in general not possible to perform the
symplectic rotation in the presence of fluxes. In particular, the rotation
exchanges a gauge boson with its magnetic dual which only is a symmetry of the
theory if no charged fields are present. The fluxes charge a subset of fields
which is another way to see the break down of the symplectic invariance.
However, if we choose $m_1^\Ah =0$ in \eqref{Db} the gauge boson in question
drops out of the covariant derivative and the symplectic rotation can be
performed.\footnote{Note that here the duals of $m_1^\Ah$ correspond to
  magnetic charges on the type IIA side. Analyzing such magnetic deformations
  of manifolds with $SU(3)$ structure goes beyond the scope of this paper and
  will be dealt with elsewhere \cite{GLW2}. For this reason only we have
  set the corresponding heterotic fluxes $m_1^\Ah$ to zero. However once these
  magnetic deformations are identified they are guaranteed to be the duals of
  $m_1^\Ah$.}

Let us now turn to the comparison of the hypermultiplet moduli space in the
two cases. This is relevant for our analysis as the charged fields
obtained by turning on fluxes on the heterotic side, and torsion on the type
IIA side, reside in hypermultiplets. Unfortunately, in this sector, the
duality is much less understood due to the more complicated structure of the
corresponding moduli spaces.\footnote{See however ref.~\cite{PA}.} For this
reason we did not consider the entire hypermultiplet sector, but
instead focused on the sub-sector of the $K3$ moduli which span the manifold
$ \mathcal{M}_{H}=SO(4,20)/SO(4) \times SO(20)$. This space is what is known to be a dual
quaternionic manifold \cite{CFG} which means that it is in the image of the
c-map or in other words it is determined by a special K\"ahler geometry. For
$\mathcal{M}_{H}$ the underlying special K\"ahler geometry is the
space \cite{CFG}
\begin{equation}
  \label{Mdual}
  \mathcal{M}_{SK} = \frac{SU(1,1)}{U(1)} \times \frac{SO(2,18)}{SO(2)
  \times SO(18)} \; .
\end{equation}

On the type IIA side we need to make a similar truncation in that out
of all hypermultiplets we only keep 20 and they are required to form the
same coset space $\mathcal{M}_{H}$ with a subspace $\mathcal{M}_{SK}$.
 The $SU(1,1)/U(1)$ factor is identified with the type
IIA dilaton/axion while the second
factor in \eqref{Mdual} has to be the moduli space of the (truncated)
complex structure 
deformations of the Calabi-Yau threefold and the associated $SU(3)$ structure
manifold respectively.  So the type IIA coordinates
of $\mathcal{M}_{SK}$ are $(\phi,a)$ and $z^a$ while the 40 RR scalars
$(\xi^A,\tilde\xi_A)$ are the coordinates which promote $\mathcal{M}_{SK}$
to $\mathcal{M}_{H}$.

On the heterotic side the coordinates of $\mathcal{M}_H$ are not unique. They are the deformations of
the three complex structures, $\vec J$, of $K3$, and the internal $B$-field.
The $SO(4)$ rotates these four fields into each other and so, there is
also
no preferred parameterization of $\mathcal{M}_{SK}$. However, turning on fluxes
breaks the $SO(4)$ invariance to $SO(3)$ since they charge the $b^\Ah$-fields.
The charged fields on the type IIA side are the $\xi^A$  which leads
us to the identification $ b^\Ah \leftrightarrow \xi^A$.
The $57$ $K3$ moduli which are the expansion coefficients of $\vec J$
are then identified with $(z^a,\tilde \xi_a)$. However, there is a
slight mismatch in that we have 22 $b^\Ah$ but only 20 $\xi^A$. 
The reason is that two `special' heterotic fluxes do not have a
geometric IIA dual of the type that we have considered in this paper.
This can be seen as follows. The heterotic--type IIA duality
constrains the $K3$ of the 
heterotic compactifications to be a $T^2$ fibred over a ${\bf P_1}$ base or in
other words the $K3$ has to be elliptically fibred. The volume of the ${\bf
  P_1}$ base on the heterotic side is identified with the type IIA
dilaton/axion. In ref.~\cite{CKKL} it was 
shown that fluxes through this ${\bf P_1}$ base, which charge the
axion with respect to all the gauge fields,
correspond to a specific set of RR fluxes on the type IIA side which
charge the dual axion again with respect to all vector fields. 
Apart from
this ${\bf P_1}$ base, the $T^2$ fibre is another special cycle among the 22 two-cycles
of $K3$. The type IIA dual of
this flux we do not understood at present  and therefore ignore it in
the following.

To summarize, only 20 cycles of $K3$ lead to fluxes which have 
a geometric IIA dual of the type considered here. For this subset of
fluxes only 20 heterotic $b$-fields 
become charged and  we denote them by $b^A$.  Then the identification above can be
refined to
\begin{equation}
  \label{map}
  \left. b^A \right|_{\mathrm{het}} \leftrightarrow \left. \xi^A
  \right|_{\mathrm{IIA}} \ .
\end{equation}
With this identification, the Killing vectors agree if we set $m^A_I = - q^A_I$ and the
fluxes through the two special cycles of $K3$ are set to zero.

Finally, let us also compare the two potentials given
\eqref{Vhet} and \eqref{VIIA}. 
First of all we note that both potentials include the respective dilatons
$\phi$ and both potentials are minimized at weak dilaton coupling
$\phi\to-\infty$. This ensures that the supergravity analysis is applicable.

Using the form of the quaternionic metric on the type IIA side, \eqref{newh},
we see that we can rewrite the first term in \eqref{VIIA} as
\begin{equation}
   \left[ - \frac{e^{2 \phi}}{4} \big(\cM \mathrm{Im} \cM^{-1} \bar \cM
      \big)_{AB} + \frac{e^{4 \phi}}{2} \txi_A \txi_B \right] 
      (\mathrm{Im} \cN^{-1})^{IJ}\, q^A_I q^B_J = -\frac12\, h_{AB} 
      (\mathrm{Im} \cN^{-1})^{IJ} q^A_I q^B_J \; ,
\end{equation}
where $h_{AB}$ denotes the restriction of the quaternionic
metric \eqref{newh} to the charged fields, $\xi^A$. This shows that the
first line in \eqref{VIIA} precisely matches the first term in the heterotic
potential \eqref{Vhet}.

In order to compare the remaining terms 
we need to use the  the special form of the hypermultiplet moduli
space $\mathcal{M}_{H}$ (and $\mathcal{M}_{SK}$) as dictated by the duality.
For the geometry \eqref{Mdual} we can use the formuli of
appendix~\ref{hetT2K3gen} where the same manifold, \eqref{Mvhet}, was
discussed for the heterotic vector multiplet sector.
Using the form of the matrix $\cM$ as given in \eqref{Nhet} 
we can  explicitely compute the first factor in the second line
of \eqref{VIIA} to be
\begin{equation}
  \label{12}
  \frac{1}{2} \big(\cM \mathrm{Im} \cM^{-1} \bar \cM \big)_{AB} + 4
  e^{K(z)} \cG_A \bar \cG_B  \sim \frac{e^{2 \phi}}{4}\, (a^2 + e^{-4 \phi})\,
  \eta_{AB} \; ,
\end{equation}
where by $\eta_{AB}$ we have denoted the invariant symmetric tensor on the
second factor of \eqref{Mdual} and $a$ denotes the type IIA axion. Similarly,
using \eqref{Nhet} again the second term in the second line of
\eqref{VIIA} yields the invariant tensor 
on the corresponding $SO(2, n_v -1)$ space
\begin{equation}
  \label{22}
  \frac12 \mathrm{Im} \cN^{-1 \, IJ} + 4 X^I \bar X^J \sim
   \frac{1}{Vol(\mathbf{P_1})}\, \eta^{IJ} \; , 
\end{equation}
where $Vol(\mathbf{P_1})$ denotes the volume of the $\mathbf{P_1}$
base of the $K3$-fibered Calabi--Yau manifold. 
In the computation of \eqref{12} and \eqref{22} we have not determined
the exact numerical factors as this would require establishing 
a precise map between the fields on both sides.
Comparing the
second line in \eqref{VIIA} with the second term in \eqref{Vhet} using
\eqref{12} and \eqref{22} we now see that they agree up to
undetermined numerical factors.

There is one final thing which can be clarified. According to what we
said above we have to identify $\rho_{AB}$ in \eqref{Vhet} with $\eta_{AB}$
in \eqref{12}. Recall that $\rho_{AB}$ is the restriction of the  $K3$
intersection matrix $\rho_{\Ah \Bh}$ where the two special cycles
defined by the ${\bf P_1}$ base  and the $T^2$ fibre are left out.
$\rho_{\Ah \Bh}$ has signature $(3,19)$
while $\eta_{AB}$ (and thus $\rho_{AB}$) has signature $(2,18)$.
The consistency can be checked as follows.
The $(3,19)$
signature of $\rho_{\Ah \Bh}$ arises from the fact that 3 of the two-forms on $K3$
are selfdual, while the other 19 are anti-selfdual. The forms which are
Poincar\'e dual to the ${\bf P_1}$ base and the $T^2$ fibre can be neither
self-
nor anti-selfdual since both cycles have zero intersection with itself,
while the product of a self-/anti-selfdual form with itself vanishes only if
the form itself vanishes.\footnote{Note that this holds for even
  (anti)self-dual forms on Euclidean spaces, which is precisely our case.}
 Let us suppose that the Poincar\'e dual form to the
base ${\bf P_1}$ has the form $S + A$ where $S$ is selfdual and $A$ is
anti-selfdual. For consistency we need that $\int S\wedge S + \int A \wedge A
=0$. Since the elliptic $T^2$ fibre is the cycle dual to the $\mathbf{P_1}$
base, its Poincar\'e dual form is the Hodge dual of $S+A$, which, by
definition is $S-A$. Therefore, removing the two cycles from our analysis is
equivalent to removing the two forms, $S$ and $A$, from the spectrum and
therefore we are left with 2 selfdual and 18 antiselfdual forms. Hence, the
restricted intersection matrix $\rho_{AB}$ has signature $(2,18)$ as it should
have been for the duality argument to work.

\section{Conclusions}
\label{conclusions}

In this paper we studied the fate of the non-perturbative duality
between heterotic string compactified on $K3\times T^2$ and
type IIA theory compactified on Calabi-Yau threefolds when fluxes on
the heterotic side are turned on. We showed that at the level of
the effective action one can restore the duality provided on the type IIA side
one considers a special class of manifolds with $SU(3)$ structure which were
proposed in \cite{GLW}. Therefore this is one of the few known examples where 
also a non-perturbative duality seems to hold in the presence of fluxes.

Another aspect of this paper is that embedding manifolds with $SU(3)$
structure satisfying \eqref{doda} into  the web string dualities is
a further (strong) argument for the existence of these manifolds.
This is important to know especially in the fields of moduli
stabilization as they induce new terms in the superpotential which can be
crucial for moduli stabilization
\cite{Ibanez,Shelton:2006fd,dCGLM,Derendinger}.

Finally let us reiterate that there are still many aspects to be better
understood in the heterotic-type IIA duality.  We already mentioned
the largely unknown structure of the hypermultiplet moduli space.\footnote{For
  recent progress in understanding the moduli space of hypermultiplets see for
  example \cite{AMTV,RSV}.}
Furthermore we only turned on a very special set of heterotic fluxes.
In \cite{LM1} it was shown
 that turning on fluxes along the heterotic $T^2$ direction 
gauges isometries in the scalar manifold of the vector
multiplets. So far such gaugings have not been discovered on the type II
side and it would be interesting to find this behavior as well. Moreover we 
have seen that on the type IIA side we have only turned on half of the
available geometric fluxes  which would mean that additional fluxes
should be possible on the heterotic side.  We hope to return to some
of these issues elsewhere.

\vspace{1cm}
\noindent
\textbf{Acknowledgments} 

We would like to thank Boris K\"ors and Andr\'e Lukas for helpful
comments and discussions. 

This work was supported in part by the European
Union 6th Framework Program MRTN-CT-2004-503369 ``Quest for Unification'' and 
MRTN-CT-2004-005104 ``ForcesUniverse''. The work of J.L.\ is also
supported by the DFG -- The German Science Foundation.

\newpage
\appendix
{\LARGE {\textbf{Appendix}}}
\vspace{1cm}

In order to make the paper self-contained we briefly review some of
the  respective properties of the $N=2$ low energy effective of 
the heterotic string compactified on $K3\times T^2$ (appendix~\ref{hetT2K3gen})
and type IIA compactified on Calabi-Yau threefolds (appendix~\ref{IIACYgen}).
We also show the consistency of $N=2$ gauged supergravity with the
heterotic string in the presence of background fluxes
(appendix~\ref{gshet}) and with type IIA compactified on
$SU(3)$-structure manifolds (appendix~\ref{gsIIA}).

\section{$N=2$  heterotic vacua in four dimensions}
\label{hetT2K3}

This appendix is devoted to the heterotic string compactified on
$K3\times T^2$. Let us first briefly review the properties of the
effective action following ref.~\cite{LM1}.

\subsection{The low energy effective action}
\label{hetT2K3gen}

The low effective action obtained from the compactifying the
heterotic string on $K3 \times T^2$ consists of a $N=2$ supergravity
coupled to $n_v$ (Abelian) vector multiplets each containing a vector, two
gaugini and a complex scalar and $n_h$ hypermultiplets each containing
two hyperions and four real scalars.
The scalars of the vector multiplets consist of the dilaton
$e^{-\phi}$, the axion $a$,
the $T^2$ moduli $G_{11}$, $G_{12}$, $\sqrt G$ and $B_{12}$,
 and the gauge fields
in the direction of the torus $A^a_1$ and $A^a_2$. 
The correct K\"ahler coordinates are
defined by (see for example \cite{LM1})
\begin{equation}
  \label{cpxdil}
  s = \frac{a}{2} - \frac{i}{2} e^{-\phi} \; ,\nn \\
\end{equation}
for the dilaton/axion while the others, $t,~u,~ n^a, ~ a = 4,\ldots , n_v$,
are given implicitly by the equations\footnote{Note that here $a$
  runs only over the Abelian gauge fields (except the KK vectors)
  which remain massless and not over the whole adjoint representation
  of the gauge group in ten dimensions.}
\begin{eqnarray}
  \label{cpxfields}
  A_1^a & = & \sqrt 2\ \frac{n^a - \bar{n}^a}{u - \bar{u}} \ ,\qquad\qquad
  A_2^a = \sqrt 2 \ \frac{\bar{u} n^a - u \bar{n}^a}{u - \bar{u}}  \ , \nn \\  
  B_{12} & = & \frac{1}{2} \left[(t + \bar{t}) - 
    \frac{(n + \bar{n})^a(n - \bar{n})^a}{u - \bar{u}} \right]\ , \nn\\
  \sqrt{G} &=&  -\frac{i}{2} \left[ 
    (t - \bar{t}) -\frac{(n - \bar{n})^a(n - \bar{n})^a}{u -
        \bar{u}} \right]\ , \\   
  G_{11} & = & \frac{2 i }{u - \bar{u}}\, \sqrt{G} \ , \qquad \qquad 
  G_{12} = i \frac{u + \bar u}{u - \bar u} \sqrt{G} \ .\nn
\end{eqnarray}
In terms of these coordinates the metric $g_{i\bar j}$ is K\"ahler, with
a K\"ahler potential 
\begin{equation}
\begin{aligned}
  \label{Kpoth}
  K &= - \, \ln i (\bar{s} -s) -\,\ln{ \frac{1}{4}\big[(t -
      \bar{t})(u - \bar{u}) - (n - \bar{n})^a  (n -\bar{n})^a  \big]} \\
  &= - \, \ln i (\bar{s} -s) - \ln{X^I \eta_{IJ} \bar X^J} \ ,
\end{aligned}
\end{equation}
where the $X^I$ denote the projective coordinates \cite{LM1,dWKLL}
\begin{equation}
  \label{Xhet}
  \begin{aligned}
    & X^0 =  \frac12 \; ,& \quad &X^1 = \frac12(ut - n^a n^a) \; , \quad X^2=
    -\frac12 u \; ,\\[2mm]
    & X^3 = \frac12 t \; ,& \quad &X^a = \frac{1}{\sqrt 2} n^a \; .
  \end{aligned}
\end{equation}
Note that the $X^I$ above satisfy
\begin{equation}
  \label{XX0}
  X^I\eta_{IJ} X^J =0 \; ,   
\end{equation}
where $\eta_{IJ}$ represents the invariant symmetric tensor of
$SO(2,n_v-1)$ which in our conventions has the form
\begin{equation}
  \label{SOmetric}
  \eta = \left(
    \begin{tabular}{ccc}
      0 & ${\bf 1}_2$ & 0 \\
      ${\bf 1}_2$ & 0 & 0 \\
      0 & 0 & ${\bf 1}$ \\
    \end{tabular} \right) \ .
\end{equation}
The geometry described above, by the K\"ahler potential \eqref{Kpoth},
corresponds to the coset manifold
\begin{equation}
  \label{Mvhet}
  \mathcal{M}_V = \frac{SU(1,1)}{U(1)} \otimes \frac{SO(2,n_v -1)}{SO(2)
  \times SO(n_v-1)} \;,
\end{equation}
where the first factor is related to the dilaton. 
As required by $N=2$ supergravity, this K\"ahler geometry is in fact a
special K\"ahler geometry. That is, the K\"ahler potential $K$ can be
expressed in terms of the quantities $(X^I, F_I)$ via
\begin{equation}
  \label{Kspec}
  K = - \, \ln i \big[\bar X^I F_I - X^I \bar F_I  \big]  \ ,
\end{equation}
where in a certain symplectic basis $F_I$ can be expressed as $F_I =
\frac{\partial F}{\partial X^I}$ with the holomorphic prepotential $F$
being a
homogeneous function of degree two in $X^I$.
It is known from ref.~\cite{Ceresole:1995jg} that $N=2$ supersymmetry only
requires that $F_I$ exists but not necessarily $F$ itself. For the
parameterization given in \eqref{Xhet} the constraint \eqref{XX0}
signals that  indeed a basis was chosen where no prepotential exists. 
In refs.~\cite{Ceresole:1995jg,dWKLL} it was shown that after the symplectic rotation $X^1\to -F_1, F_1\to X^1$ the
K\"ahler potential can be derived from the prepotential 
\begin{equation}
  \label{Fhet}
  F\ =\ \frac{X^1(X^2 X^3 - X^aX^a)}{X^0}\ .
\end{equation}

The other couplings of the vector multiplets which appear in the
action \eqref{s4het} are the gauge couplings $\cN$.
They are given by 
\begin{equation}
  \label{Nhet}
  \cN_{IJ} = - \frac{s + \bar s}{2} \eta_{IJ} + \frac{s- \bar s}{2}
  \left(\eta_{IJ} - 2 \frac{(X_I \bar X_J + \bar X_I X_J)}{X^K
  \eta_{KL} \bar X^L} \right)\ ,
\end{equation}
where the indices on $X$ are raised and lowered with the metric $\eta$
defined in \eqref{SOmetric}. 

Finally, the couplings of the hypermultiplets are encoded in the
quaternionic metric $h_{uv}$ of the
action \eqref{s4het}. The scalar fields of the  hypermultiplets
consist of the $K3$ moduli which span the coset
\begin{equation}
  \label{Mhhet}
  \mathcal{M}_H =\frac{SO(4,20)}{SO(4) \times SO(20)}\ ,
\end{equation}
 together with the bundle moduli whose
number and couplings can be determined only once a specific solution
to \eqref{Bid} has been chosen. In the absence of a concrete solution to
\eqref{Bid} the details of the hypermultiplet moduli space are not
known. In the main text we mainly   concentrate on the ``model
independent'' part of the metric which is the metric on the
moduli space of $K3$.

\subsection{Consistency with gauged supergravity}
\label{gshet}

The purpose of this appendix is to show how the potential $V$ given in
\eqref{Vhet} together with the covariant derivatives \eqref{Db} is
consistent with the general constraints of gauged $N=2$ supergravity
as for example given in ref.~\cite{N2}. In fact most of this proof is
already contained in ref.~\cite{LM1} where the
consistency for $\fx=0$ was shown. Here we do not want
to repeat all the details but merely argue how the proof of \cite{LM1}
carries over to the case $\fx \neq0$.

The main point is that the covariant derivatives \eqref{Db} do not depend on
the value of $\fx$ and therefore are unchanged. This in turn leaves the
Killing vectors $k_I^u$ defined by $Dq^u = \partial q^u -k_I^uA^I$ unchanged.
From \eqref{Db} we infer
\begin{equation}
  \label{Killing}
k_I^u = m^\Ah_I  \; .
\end{equation}
In $N=2$ supergravity the Killing vectors are expressed in
terms of a triplet of Killing prepotentials $ \cP^x_I, x=1,2,3$
defined by $k_I^u K^x_{uv} = - D_v  \cP^x_I$, where $K^x_{uv}$ is a
triplet of almost complex structures which always exists on a
quaternionic manifold. Since $k_I^u$ is unchanged for $\delta\neq0$
also the $\cP^x_I$
remain unchanged. The computation of the $\cP^x_I$ we do not repeat
here but just take them from appendix~D of ref.~\cite{LM1}. More precisely
let us recall the result\footnote{Note
  that due to a convention mismatch the second term in this
  equation came with a minus sign in \cite{LM1}. However, for
$\delta=0$ this term did not contribute to the potential and therefore
its sign was not important in \cite{LM1}.}
\begin{equation}
  \label{pxpx}
  \cP^x_I \cP^x_J = h_{\Ah \Bh} m^\Ah_I m^\Bh_J + \frac{\rho_{\Ah \Bh} m^\Ah_I
  m^\Bh_J}{4v} \; . 
\end{equation}

If only isometries in the hypermultiplet sector
are gauged the $N=2$ scalar potential reads \cite{N2}
\begin{equation}
  \label{VN2}
  V_{N=2} = 4e^K X^I \bar X^J h_{uv}k_I^uk_J^v
  - \left(\frac12\, (\mathrm{Im} \cN^{-1})^{IJ}  + 4e^K X^I \bar X^J\right)
  \cP^x_I \cP^x_J \; . 
\end{equation}
Inserting \eqref{Killing} and \eqref{pxpx} yields
\begin{equation}
  \label{VN2h}
  V_{N=2} = - \frac12\, (\mathrm{Im} \cN^{-1})^{IJ} h_{\Ah \Bh} m^\Ah_I
  m^\Bh_J - \left( \frac12  (\mathrm{Im} \cN^{-1})^{IJ} + 4 e^K X^I
  {\bar X}^J  \right)\,  \frac{\rho_{\Ah \Bh} m^\Ah_I m^\Bh_J}{4v}\ .
\end{equation}
Using the equations \eqref{Kpoth}--\eqref{Nhet} one finds
\begin{equation}
  \label{N-1het}
 (\mathrm{Im} \cN^{-1})^{IJ} = \frac{2i}{s -\bar s}\, \eta^{IJ} 
  - 4 e^{K} (X^I  \bar X^J + \bar X^I X^J) \; .
\end{equation}
Inserted into \eqref{VN2h} and using the identity $X^I \eta_{IJ} X^J
=0$  we finally arrive at 
\begin{equation}\label{VN2hf}
  V_{N=2} = - \frac12 h_{\Ah \Bh} m^\Ah_I m^\Bh_J (\mathrm{Im} \cN^{-1})^{IJ} +
   \frac{e^\phi}{4v} \; m^\Ah_I m^\Bh_J \rho_{\Ah \Bh } \eta^{IJ} \; ,
\end{equation}
which indeed coincides with the potential \eqref{Vhet} derived from
the compactification.  This establishes the consistency with $N=2$
gauged supergravity. The second term in \eqref{VN2hf} was missing in
\cite{LM1} due to the constraint $\delta=0$.

\section{$N=2$ type IIA vacua in four dimensions}
\label{IIACY}

\subsection{The low energy effective action}
\label{IIACYgen}
We now briefly recall the structure of the $N=2$ supergravity obtained
from Calabi--Yau compactifications of type IIA string theory which was
first discussed in \cite{BCF}. (Our presentation here however follows more closely ref.~\cite{LM2}.) 

One starts from the ten-dimensional action \eqref{SA10} and expands
the ten-dimensional fields in terms of the harmonic forms on the
Calabi--Yau manifold $Y$. These are the $h^{(1,1)}$ $(1,1)$-forms 
$\ox_i,\, i=1, \ldots, h^{(1,1)}$ and the harmonic three-forms 
$(\alpha_A, \beta^A)$, where $A=0, \ldots , h^{(2,1)}$.
There also is  a set of dual four-forms $\tox^i$ 
 such that
\begin{equation}
  \label{norm2}
  \int_Y \ox_i \wedge \tox^j = \delta_i^j  \; .
\end{equation}
Similarly, the harmonic three-forms can be chosen to form a
symplectic basis of the third cohomology group $H^3$  such that
\begin{equation}
  \label{norm3}
  \int_Y \alpha_A \wedge \beta^B = \delta_A^B \; , \qquad \int_Y
  \alpha_A \wedge \alpha_B = \int_Y \beta^A \wedge \beta^B = 0 \; .
\end{equation}

The ten-dimensional massless fields $\hat B_2, \hat C_1, \hat C_3$ are
then Kaluza-Klein expanded according to
\begin{eqnarray}
  \label{fexpCY}
  \hat B_2 & = & B_2 + b^i \omega_i \; , \nn \\
  \hat C_1 & = & A^0 \; , \\
  \hat C_3 & = & C_3 + A^i \wedge \omega_i +  \xi^A \alpha_A - \tilde \xi_A
  \beta^A \; , \nn
\end{eqnarray}
where $B_2$ is a two form, $(A^0,A^i)$ are one-forms and $b^i,
\xi^A,\tilde\xi_A$ are scalars.
Furthermore the Calabi-Yau metric has two sets of independent
deformations, the deformations $v^i$, of the K\"ahler form $J$, and the
deformations $z^a$, of the complex structure, which can  equivalently be
viewed as the deformations of
the holomorphic three-form $\Omega$
\begin{equation}
  \label{JOexpCY}
  \begin{aligned}
      J = v^i \omega_i \; ,  \qquad
      \Omega =  Z^A \alpha_A - \cG_A \beta^A \; .
  \end{aligned}
\end{equation}
$\cG_A = \frac{\partial \cG}{\partial Z^A}$ is the derivative of the
holomorphic prepotential $\cG$ and the complex structure moduli are given by
$z^a = Z^a/Z^0$. Altogether, these fields combine into 
$h^{(1,1)}$ vector multiplets, consisting of the bosonic components
$(A^i, x^i = b^i + iv^i)$ and $h^{(2,1)}$ hypermultiplets consisting of the
scalars $(z^a, \xi^a, \txi_a)$. In addition there is a tensor
multiplet with components $(B_2, \phi, \xi^0, \txi_0)$ and finally
$A^0$ is the graviphoton which sits in the $N=2$ gravitational
multiplet. 

If the tensor multiplet is dualized to an additional hypermultiplet,
the effective action is again of the standard $N=2$ form as given in
\eqref{s4het}. The metric  $g_{ij}$ of the scalars in the
vector multiplets is defined as
\begin{equation}
  \label{gKd}
  g_{ij} = \frac1{4 \cK} \int_Y \ox_i \wedge * \ox_j \; ,
\end{equation}
where $\cK$ is the volume of $Y$. 
As required by $N=2$ supergravity this metric is special K\"ahler and 
can be derived from the prepotential
\begin{equation}
  \label{FIIA}
  F = - \frac16 \cK_{ijk} x^i x^j x^k \; ,
\end{equation}
where $\cK_{ijk}$ denotes the triple intersection numbers of the
Calabi--Yau manifold.
Furthermore, the imaginary part of the gauge coupling matrix $\cN$
in \eqref{s4het} is given by\footnote{The real part of this matrix 
plays no role for the analysis here but can be found, for example, in
ref.~\cite{LM2}.}  
\begin{equation}
  \label{N}
  \mathrm{Im} \cN = - \cK\left(
    \begin{array}{cc}
      1 + 4 g_{ij} b^i b^j & 4 g_{ij} b^j \\
      4 g_{ij} b^i & 4 g_{ij}
    \end{array}
    \right) \; .
\end{equation}

On the type IIA side the  quaternionic metric $h_{uv}$ on the space of
hypermultiplet scalars can be given explicitly.
It is determined in terms of the 
special K\"ahler geometry which describes the complex
structure deformations $z^a$ on a Calabi--Yau manifold. One first
defines the matrix $\cM_{AB}$ by the following integrals\footnote{Note
  that due to the Hodge $*$ these integrals depend on the choice of
  complex structure or in other words on $z^a$.}
\begin{equation}
  \label{M}
  \begin{aligned}
    \int_Y \alpha_A \wedge * \alpha_B = & \; - \mathrm{Im} \cM_{AB} -
    \mathrm{Re} \cM_{AC} (\mathrm{Im} \cM)^{-1 \; CD} \mathrm{Re}
    \cM_{DB}  \; , \\
    \int_Y \alpha_A \wedge * \beta^B = & \; - \mathrm{Re} \cM_{AC}
    (\mathrm{Im} \cM)^{-1 \; CB} \; , \\
    \int_Y \beta^A \wedge *\beta^B = & \; - (\mathrm{Im} \cM)^{-1 \; AB} \; .
  \end{aligned}
\end{equation}
The quaternionic metric is then expressed in terms of $\cM_{AB}$ by \cite{FeS}
\begin{eqnarray}
  \label{qkt}
  h_{uv} Dq^u \wedge * Dq ^v & = & \frac14 (d\phi)^2 + g_{a \bar b} dz^a
  \wedge * d \bar z^b + \frac{e^{4\phi}}{4} \, \Big[Da - (\tilde\xi_A D
  \xi^A - \xi^A D \tilde \xi_A )\Big]^2  \\
  & & - \frac{e^{2\phi}}{2}\left(\mathrm{Im} \cM^{-1} \right)^{AB} 
  \Big[ D\tilde\xi_A - \cM_{AC} D\xi^C \Big]
   \wedge * \Big[ D\tilde\xi_B - \bar \cM_{BD} D\xi^D \Big]  \, .\nn
\end{eqnarray}
where, $g_{a \bar b}$ is the metric on the complex structure moduli
space and $a$ denotes the (pseudo)-scalar which is dual to the
two-form $B_2$ in four dimensions.

\subsection{Consistency with gauged supergravity}
\label{gsIIA}

The purpose of this appendix is to show the consistency of the
potential \eqref{VIIA} with the general form of the $N=2$ potential
\eqref{VN2}. Note that this can be written as
\begin{equation}
  \label{VN2alt}
  V_{N=2} = \left[\frac12 (\mathrm{Im} \cN^{-1})^{IJ} + 4 e^{K(X)} X^I \bar
  X^J \right] (h_{uv} k^u_I k^v_J - \cP^x_I \cP^x_J ) - \frac12 (\mathrm{Im}
  \cN^{-1})^{IJ} k^u_I k^v_J h_{uv} \ .
\end{equation}
As in the heterotic case we first need to determine the
Killing prepotentials $\cP^x_I$. For generic manifolds with $SU(3)$
structure they were computed in ref.~\cite{GLW}. For the
case at hand they read
\begin{equation}
\begin{aligned}
\cP^1_I + i\cP^2_I = 2e^{\tfrac12 K(z) + \phi} q_I^A \cG_A\ , \qquad
\cP^3_I &= e^{2\phi} q_I^A \tilde\xi_A\ .
\end{aligned}
\end{equation}
This enables us to also compute
\begin{eqnarray}
  \label{Ps}
  \cP^x_I \cP^x_J = 4 e^{K(z)} e^{2 \phi} \cG_A \bar \cG_B q^A_I q^B_J + e^{4
  \phi} \txi_A \txi_B q^A_I q^B_J \; .
\end{eqnarray}
Inserted into \eqref{VN2alt} and also using $k^u_I = -q_I^A$ as can be seen
from 
\eqref{cdxi} and $h_{uv}$ given in \eqref{newh} it is now not hard to see that
the above potential precisely coincides with the one obtained in \eqref{VIIA}.
This establishes that the result on the type IIA side does indeed describe a
$N=2$ gauged supergravity.

\end{document}